\documentclass[10pt,journal,twoside]{IEEEtran}
\IEEEoverridecommandlockouts

\usepackage{latexsym,epsf,graphicx,psfrag,amssymb,cite}
\usepackage{cite,tabularx}
\usepackage{url}
\usepackage{epsf}
\usepackage{amsmath,tikz}
\usepackage{amssymb}
\usepackage{pifont}
\usepackage{epsfig}
\usepackage{psfrag}
\usepackage{graphicx}
\usepackage{subfigure}
\usepackage{multirow}
\usepackage{array}
\usepackage{enumerate}
\usepackage{algorithm}
\usepackage{algorithmic}
\usepackage{amsthm}
\usepackage{subfigure}
\usepackage{float,lscape}
\usepackage{longtable}
\usepackage{caption}
\usepackage{pdflscape}
\usepackage{rotating}
\usepackage{ragged2e}
\usepackage{soul}
\usepackage[acronym]{glossaries}
\usepackage[normalem]{ulem}
\newcommand{\cmark}{\ding{51}}%
\newcommand{\xmark}{\ding{55}}%
\usepackage{lineno}

\def\BibTeX{{\rm B\kern-.05em{\sc i\kern-.025em b}\kern-.08em
    T\kern-.1667em\lower.7ex\hbox{E}\kern-.125emX}}

\begin{document}

\title{Resource Optimization in UAV-assisted IoT Networks: The Role of Generative AI}

\author{Sana~Sharif, Sherali~Zeadally, and Waleed~Ejaz
\thanks{(Corresponding author: Waleed Ejaz.)

S. Sharif and W. Ejaz is with the Department of Electrical and Computer Engineering, Lakehead University, ON, Canada (email: ssharif@lakeheadu.ca; wejaz@lakeheadu.ca).\\

S. Zeadally is with the College of Communication and Information, University of Kentucky, Lexington, KY, USA (email: szeadally@uky.edu). \\

}}

\maketitle

\begin{abstract}
We investigate how generative Artificial Intelligence (AI) can be used to optimize resources in Unmanned Aerial Vehicle (UAV)-assisted Internet of Things (IoT) networks. In particular, generative AI models for real-time decision-making have been used in public safety scenarios. This work describes how generative AI models can improve resource management within UAV-assisted networks. Furthermore, this work presents generative AI in UAV-assisted networks to demonstrate its practical applications and highlight its broader capabilities. We demonstrate a real-life case study for public safety, demonstrating how generative AI can enhance real-time decision-making and improve training datasets. By leveraging generative AI in UAV-assisted networks, we can design more intelligent, adaptive, and efficient ecosystems to meet the evolving demands of wireless networks and diverse applications. Finally, we discuss challenges and future research directions associated with generative AI for resource optimization in UAV-assisted networks.
\end{abstract}

\section{Introduction}
\label{section:Introduction}
Unmanned Aerial Vehicle (UAV)-assisted networks have emerged as a transformative solution to enhance connectivity in Internet of Things (IoT) systems. These IoT networks utilize the mobility and versatility of UAVs to offer dynamic coverage extension and on-demand deployment capabilities, which are particularly valuable in remote or challenging environments. UAVs can serve as aerial base stations, relays, or mobile gateways, effectively bridging communication gaps and enabling real-time data exchange in diverse IoT applications such as disaster management, agriculture, and infrastructure monitoring \cite{2-mcenroe2022survey}. However, integrating UAVs into IoT ecosystems presents unique challenges, including technical complexities, regulatory constraints, and operational considerations. Issues such as aerial mobility, energy efficiency, airspace regulations, and privacy concerns require careful attention to ensure UAV-assisted networks' safe and effective deployment. Addressing these challenges requires collaborative efforts across multiple disciplines, ranging from wireless communication and robotics to policy-making and regulatory compliance, to unlock the full potential of aerial networking in IoT systems.

Resource optimization refers to efficiently allocating and utilizing available resources to achieve the desired objectives or maximize performance in a given system \cite{40-9501056}. Resource optimization is crucial in ensuring the effective operation and sustainability of interconnected environments in the context of UAV-assisted IoT networks. With the proliferation of IoT devices and the integration of UAVs into communication infrastructures, resource optimization becomes increasingly critical to address challenges such as limited bandwidth, energy constraints, and dynamic network conditions. This can be done by optimizing spectrum, power, computing resources, and flight trajectories. UAV-assisted networks can enhance connectivity, minimize latency, and maximize the quality of service for IoT applications across diverse domains. The importance of resource optimization in UAV-assisted IoT networks arises from the complex interplay among various factors such as mobility, scalability, and heterogeneous communication technologies. Efficient resource optimization is essential to meet the diverse requirements of IoT devices and applications, ranging from low-latency data transmission to energy-efficient operation. 

\begin{figure}
    \centering
    \includegraphics[width=.50\textwidth, height= 8cm]{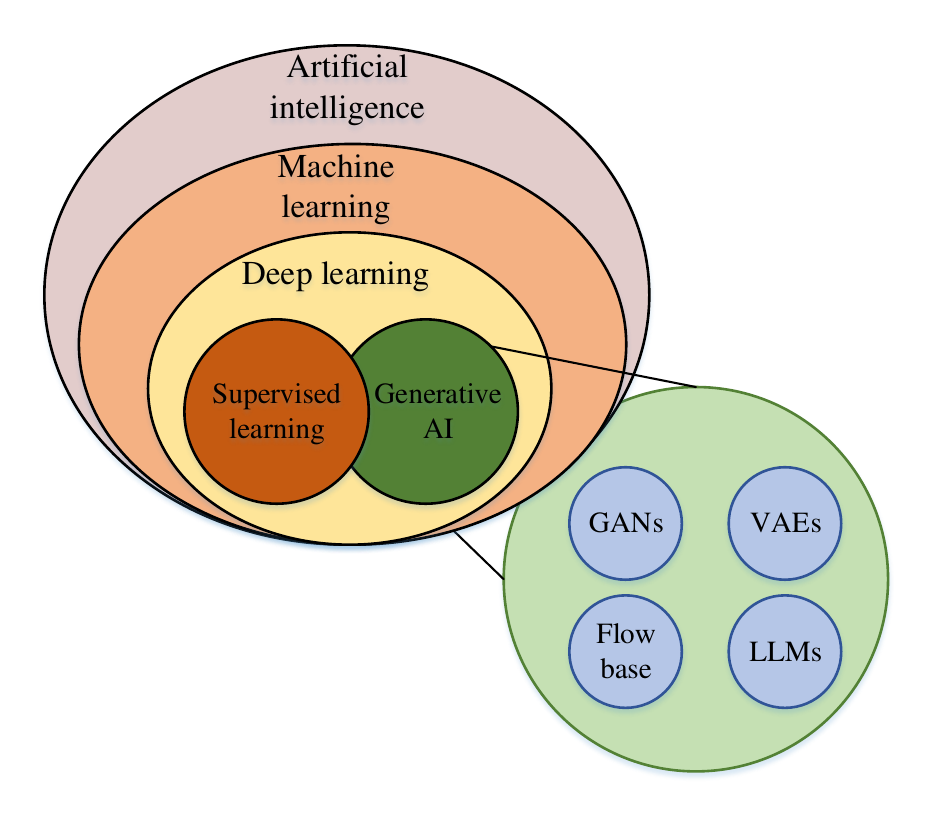}
    \caption{AI and Generative AI models. GAN: Generative Adversarial Networks; VAEs: Variational AutoEncoders; LLMs: Large Language Models.}
    \label{fig:model}
\end{figure}

Fig. \ref{fig:model} shows generative artificial intelligence (AI), a subset of AI that involves algorithms and models designed to generate new data instances or content that resemble real-world samples. These algorithms use advanced techniques such as neural networks and probabilistic frameworks to learn the underlying patterns and structures of data, enabling them to produce novel outputs with diverse applications. In UAV-assisted IoT networks, generative AI holds tremendous potential for addressing resource optimization challenges \cite{13-wang2023unified}. By leveraging generative models, UAVs can dynamically generate optimal flight trajectories, allocate spectrum resources efficiently, and optimize energy consumption, thereby enhancing the performance and scalability of IoT networks. Generative AI can learn complex relationships and patterns from data, making informed decisions and adapting to changing environmental conditions in real-time. Generative models can analyze vast amounts of historical and real-time data to predict future trends, identify optimization opportunities, and generate resource allocation and management strategies. Moreover, generative AI models such as Generative Adversarial Networks (GANs) and Variational Auto Encoders (VAEs) can generate synthetic data to augment training datasets, facilitate simulation-based optimization, and enhance the robustness and generalization of resource optimization algorithms in UAV-assisted IoT networks.

We explore how generative AI can be used to improve the performance and scalability of UAV-assisted IoT networks, leading to safer, more efficient, and resilient IoT ecosystems. By investigating the practical applications and benefits of generative AI models in UAV-assisted networks, we demonstrate the potential to enhance real-time decision-making, improve training datasets, and create more intelligent and adaptive ecosystems to meet the evolving demands of wireless networks and diverse IoT applications.

We explore the benefits of generative AI for resource optimization in UAV-assisted IoT networks. The main research contributions of this paper are:
\begin{itemize}
    \item We explore the potential of generative AI to optimize resource allocation in UAV-assisted IoT networks.
  \item We examine the impact of generative AI models on resource optimization for UAV-assisted IoT networks.
    \item We present a real-world use case in public safety to demonstrate the practical application of generative AI in enhancing resource management and real-time decision-making in UAV-assisted IoT networks.
    \item We identify challenges and highlight future research directions and potential solutions to meet real-time requirements in UAV-assisted IoT networks.    
\end{itemize}

\begin{figure*} [t]
\centerline{
\includegraphics[width=6.5in, trim = 0.5cm 5cm 0.5cm 5cm,clip]{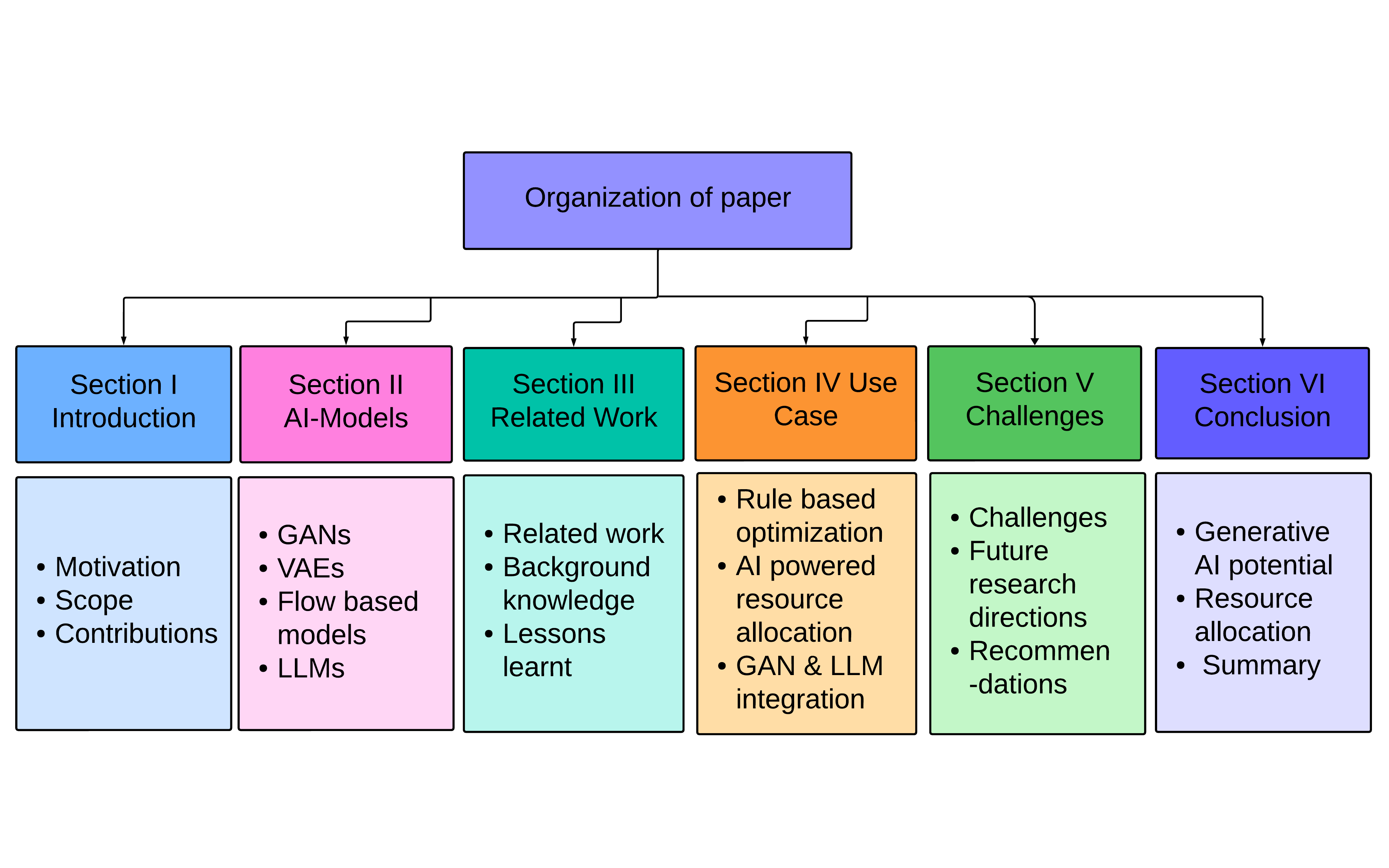}
}
\caption{Organization of the paper.}
\label{fig:organization}
\end{figure*}

Fig. \ref{fig:organization} describes the paper's organization.

\section{Generative AI Models}
Various generative AI models, including GANs, energy-based models, Variational AutoEncoders (VAEs), flow-based models, and diffusion models, aim to replicate actual data distributions through iterative training \cite{22-bond2021deep}. Each model has a unique architecture and target applications.

\subsection{Generative Adversarial Networks (GANs)}  
GANs represent a category of machine learning algorithms designed to generate authentic-looking synthetic data, as Fig. \ref{fig:models}(a) shows. These networks comprise two key components: a generator, responsible for crafting counterfeit data (X'), and a discriminator, tasked with differentiating between genuine (X) and fabricated data (X'). Through an adversarial training process, the generator and discriminator engage in a competitive learning dynamic learning process. 
Over time, the generator refines its ability to produce synthetic data that resemble real ones, creating a highly realistic output. At the same time, data features are represented with (Z). This adversarial setup defines the essence of GANs \cite{2-mcenroe2022survey}.

\subsection{Variational Autoencoders (VAEs)} 
VAEs use pairs of encoder and decoder networks to learn patterns in data without needing labelled information, as Fig. \ref{fig:models}(b) shows. This approach is distinct from GANs because VAEs focus on a process where the network learns to encode input data unsupervised rather than generating data through the adversarial competition like GANs need \cite{22-bond2021deep}.

\subsection{Flow Based Models}
\begin{figure*}
\centering
\subfigure[]{\includegraphics[width=3.5in, trim = 0.1cm 0.1cm 0.1cm 0.1cm,clip]{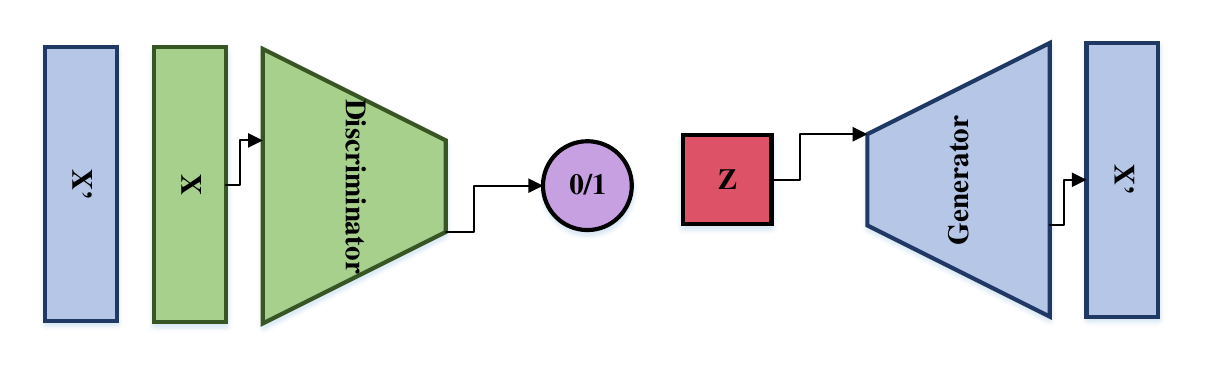}}
\subfigure[]{\includegraphics[width=3in, trim = 0.1cm 0.1cm 0.1cm 0.1cm,clip]{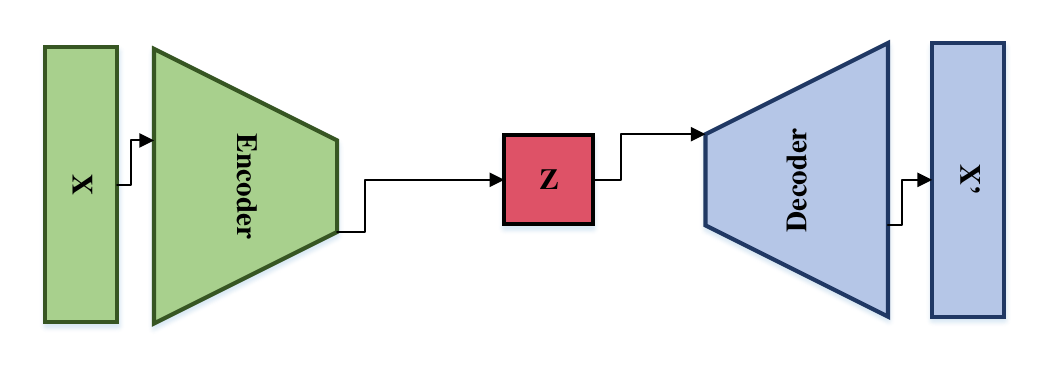}}
 \subfigure[]{\includegraphics[width=3.5 in, trim = 0.1cm 0.1cm 0.1cm 0.1cm,clip]{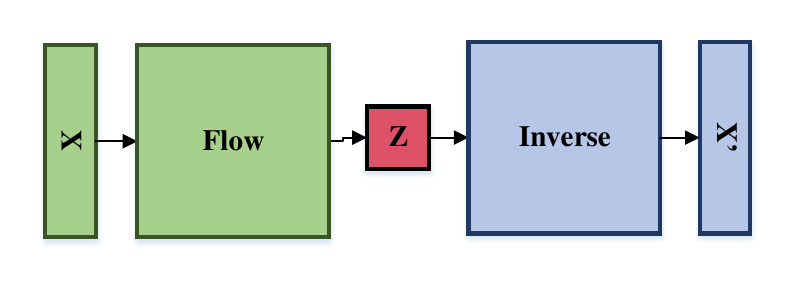}}
 \subfigure[]{\includegraphics[width=3.5 in, trim = 0.1cm 0.1cm 0.1cm 0.1cm,clip]{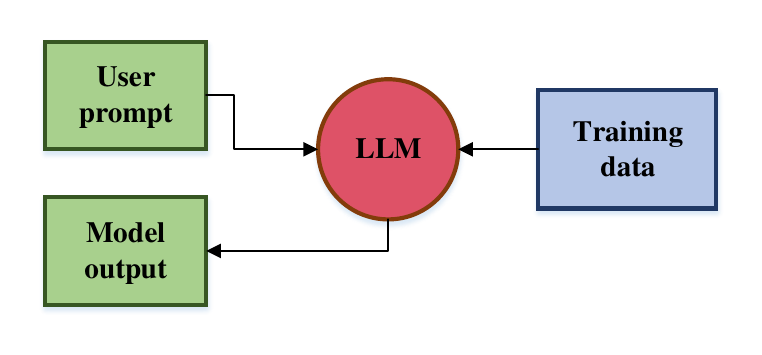}}
\hspace{-2em}
\caption{Generative AI models (a) Generative adversarial networks, (b) Variational autoencoder, (c) Flow based model, and (d) Large language model.}
\label{fig:models}
\end{figure*}

Flow-based models use specific mathematical formulations related to probability to generate data efficiently. This is especially advantageous in mobile edge networks, where creating data efficiently is crucial. A sequence of invertible transformations constructs a flow-based generative model, as Fig. \ref{fig:models}(c) shows. Unlike GANs and VAEs, the flow-based model explicitly learns the data distribution; therefore, the loss function is simply the negative log-likelihood\cite{24-xu2023unleashing}. 

\subsection{Large Language Models (LLMs)} 
Fig. \ref{fig:models}(d) shows the concept of Large Language Models (LLMs), which are advanced AI models designed to understand and generate human-like language. LLMs are characterized by an extensive number of parameters, often in the billions. Examples of such models include OpenAI's GPT-3 (generative pre-trained transformer 3) and similar models.  These models are trained on vast amounts of diverse text data, enabling them to perform various natural language processing tasks, such as text completion, translation, question-answering, and more. However, LLMs face challenges in wireless networks, requiring efficient resource utilization and compatibility with existing infrastructures\cite{46-kasneci2023chatgpt}.

LLMs deployed in UAV-mobile edge computing can address issues such as long response times, high bandwidth requirements, and data privacy concerns. However, practical caching algorithms must balance latency, energy consumption, and accuracy in edge servers. This work aims to enable UAV networks to meet the dynamic demands of UAV assisted IoT networks 
 Overall, generative AI has enormous potential to enhance resource optimization and efficiency in UAV-assisted IoT networks.

\section{Related Works}
Generative AI has strong potential in UAV-assisted IoT networks through its ability to optimize the utilization of resources. It can efficiently manage bandwidth, prioritize critical data transmission, select the most suitable payload, compress sensor data, dynamically allocate resources, enhance spectrum efficiency, facilitate autonomous decision-making, predict maintenance needs, augment real-world data with synthetic data for machine learning, simulate network scenarios for training, and supporting data analytics. Current research explores the role of generative AI in optimizing UAV trajectory planning, resource allocation, and connectivity. It addresses energy efficiency challenges, ensures attack-resilient connectivity, and guides digital content generation services. Using generative AI in UAV-assisted IoT networks demonstrates its potential to revolutionize resource optimization, adaptability, and intelligence in dynamic environments.

 
 Moreover, generative AI enables UAVs to make autonomous decisions regarding resource allocation and utilization there by minimizing downtime through predictive maintenance and facilitating data-driven tasks by augmenting real-world data with synthetic data. Leveraging generative AI for realistic simulations and data analytics further enhances operators' resource management strategies, ultimately advancing the capabilities and effectiveness of UAV systems in diverse operational scenarios.

The authors of \cite{1-bilgram2023accelerating} discussed generative AI LLMs (e.g., GPT) in innovation management, exploration, ideation, and digital prototyping. It highlights how AI can enhance innovation processes and change how companies work and innovate in different areas, which can also potentially manage UAV-assisted IoT networks. The authors presented use cases, including AI-augmented digital prototyping, which enables non-technical individuals to create prototypes by describing them in natural language. The authors of \cite{2-mcenroe2022survey} have proposed an innovative approach that integrates edge computing and AI for UAVs. The authors discussed the impact of edge AI on various technical aspects of UAVs, and these include autonomous navigation, controlling the formation of UAV clusters, power management, security \cite{44-inayat2022learning} and privacy, computer vision, and communication \cite{45-rafique2020communication}.


The authors of this survey \cite{39-karapantelakis2023generative} studied  5th Generation (5G) networks utilizing discriminative AI algorithms for network analysis, optimization, and troubleshooting tasks. These algorithms rely on historical or real-time network data for machine learning model training. With the emergence of 6th Generation (6G) which aims to meet more demanding connectivity requirements, challenges arise regarding data observability and safe learning. Generative AI models offer solutions by providing data for training in cases where data collection is resource-intensive or when online learning approaches pose risks of inaccurate predictions. Additionally, as customer-related tasks become more complex in 6G networks, generative AI solutions, including transformers and natural language processing, can aid in eliciting user requirements while addressing privacy concerns. These challenges extend beyond 6G to existing mobile networks, necessitating ongoing research efforts for practical solutions.

This paper \cite{40-9501056} introduced a novel framework for modeling air-to-ground channels in millimeter wave (mmWave) frequencies within a UAV wireless network. The approach developed an effective channel estimation method, allowing each UAV to create a local channel model through a GAN. The authors designed a cooperative framework based on a distributed GAN architecture to enable privacy-preserving sharing of channel information among UAVs. The paper established the necessary and sufficient conditions for an optimal network structure that maximizes the learning rate during information sharing. Simulation results showed that the proposed GAN approach exhibits an increased learning rate by sharing more generated channel samples in each iteration, albeit with a decrease in additional UAVs. Furthermore, the GAN method demonstrated higher learning accuracy than a standalone GAN, resulting in a more than 10\% improvement in the average UAV downlink communications rate compared with a baseline real-time channel estimation scheme.

This article \cite{41-9771337} investigated the utilization of AI in addressing energy efficiency challenges in UAV-assisted wireless networks, particularly in the context of the 6G wireless network. The focus is on enhancing energy efficiency in three key areas: trajectory planning for UAVs, resource allocation encompassing caching, computing, and communication for UAVs, and decision-making for the 3D hovering locations of UAVs. The authors discussed each aspect, highlighting the optimization challenges related to energy efficiency. 
To address these optimization challenges, the article explored several promising AI methods based on deep learning, including pointer networks, federated deep learning, and multi-agent deep deterministic policy gradients. The case studies presented validated the effectiveness of the proposed AI methods in conserving UAV energy and reducing system delays.


Similarly, in \cite{10-zou2023wireless}, the authors explored the promising technique of generative LLMs, edge networks, and multi-agent systems for wireless networks. The authors introduced the concept of on-device LLMs and their collaborative role in achieving network goals (such as network energy saving and user transmission rate), emphasizing the limitations of cloud-based LLMs and discussing multi-agent LLMs from a game theoretic perspective. The article also investigated the architectural design for wireless multi-agent generative AI systems and identified vital wireless technologies (such as on-device LLMs). The authors presented a case study highlighting the advantages of wireless productive agents in intent-based networking and outlining potential challenges (such as resource management for on-device LLMs). This research on generative LLMs, edge networks, and multi-agent systems in wireless networks is pertinent to generative AI and UAV-assisted networks, demonstrating their collaborative potential for efficient and adaptive communication infrastructures in dynamic environments.


The authors of \cite{11-raha2023generative} presented a novel semantic communication framework for next-generation wireless networks to address the challenges of excessive data transmission, which leads to high bandwidth consumption, increased latency, and reduced quality of services for applications like Intelligent Transportation Systems (ITS), metaverse, mixed reality, and the Internet of everything. The framework includes an intelligent semantic transmitter to capture meaningful information efficiently and a GAN-based semantic communication for data reconstruction and denoising, and it has been tested in real-world 6G ITS scenarios. The results show a significant 93.45\% reduction in data communication while maintaining high-quality data reconstruction across various signal-to-noise channel conditions. This work showcases the pivotal role of generative AI in optimizing communication efficiency for emerging applications such as ITS, metaverse, mixed reality, and the Internet of Everything, which can also include UAV-assisted IoT networks.

In \cite{12-du2023generative}, the authors discussed the challenges of efficiently deploying AI-Generated Content (AIGC) models for the emerging metaverse paradigm by introducing an AIGC-as-a-service architecture in wireless edge networks. It selects AIGC service providers efficiently considering environmental uncertainty and variability, and proposed the diffusion
soft actor-critic algorithm, which combines deep reinforcement learning and a diffusion model-based AI-generated optimal decision algorithm. Extensive experiments demonstrate diffusion soft actor-critic's superiority over seven other deep reinforcement learning algorithms, highlighting its potential for optimizing AIGC-driven services in the Metaverse. This work explored the significance of generative AI in optimizing decision-making processes, especially crucial in UAV-assisted IoT networks where environmental uncertainty and variability could be 
addressed through the advanced diffusion soft actor-critic algorithm, emphasizing its potential for enhancing AIGC-driven services. 

The authors of \cite{13-wang2023unified} proposed a framework that utilized wireless perception (the ability of a system to interpret and make sense of information from the environment)
to guide generative AI for digital content generation services in resource-constrained mobile edge networks. It introduced a sequential multi-scale perception algorithm to predict user skeletons radio-frequency identification can also be used for human skeleton\cite{13-wang2023unified }
estimation
 based on channel state information from wireless signals, guiding generative AI in generating AIGC. The authors developed a pricing-based incentive mechanism and a diffusion model-based approach is designed to ensure efficient operation to optimize the pricing strategy for service provisioning. Experimental results demonstrated the framework's effectiveness in skeleton prediction and optimal pricing strategy generation compared to existing solutions. This work on pricing strategy can also be used in UAV-assisted IoT networks as UAVs can be transferred to crowded areas where more coverage is required.

In \cite{14-yang2022generative}, the authors discussed the significance of trust in the emerging 6G wireless networks and emphasized the role of AI in achieving trustworthy services. The authors reviewed AI-based trust management schemes and proposed a generative-adversarial-learning-enabled trust management method for 6G and outlined a heterogeneous and intelligent 6G architecture. The authors integrated AI and trust management to enhance intelligence and security. The proposed AI-based trust management method to secure clustering for reliable and real-time communications is applied, with simulation results confirming its excellent performance in ensuring network security and service quality. This work emphasized the synergy between generative AI and trust management to enhance intelligence and security, offering insights into the applicability of generative AI in ensuring trustworthy services in the context of UAV-assisted IoT networks.

\noindent \textbf{Lessons learnt from related works:}
Table I presents a comprehensive summary of current research on generative AI. The research offers insights into the significant developments, challenges, and opportunities of generative AI in UAV-assisted IoT networks. Generative AI, particularly GANs and LLMs, is widely used to improve UAV-assisted IoT network operations. These AI models optimize resource utilization, enhance adaptability, and facilitate autonomous decision-making in dynamic environments. The reviewed works focus on energy efficiency, edge computing, and data rate optimization, emphasizing their significance in ensuring efficient UAV operations. Additionally, integrating generative AI with edge computing technologies improves various technical aspects of UAV systems, such as autonomous navigation and power management.  Overall, exploring generative AI is a potential approach for revolutionizing resource optimization, intelligence, and security in UAV-assisted IoT networks, significantly enhancing operational capabilities.



\begin{table*} 
\centering
\caption{Related work on generative AI for UAV-assisted IoT networks (\textbf{G} = Generative AI and \textbf{U}= UAV assisted network).}
\label{Existing-Survey}
\begin{tabular}{|p{0.25 in}|p{0.25in}|p{0.25in}|p{0.75in}|p{0.25 in}|p{1.5in}|p{2.5 in}|}
\hline
 \textbf{Ref.} & \textbf{Year} & \textbf{G}&\textbf{Generative AI model}& \textbf{U} &\textbf{Focus area}  & \textbf{Remarks} \\ \hline

    \cite{40-9501056} & 2021  &\cmark &  GANs &\cmark & A practical channel estimation approach is developed to collect mmWave channel information, allowing each UAV to train a local channel model via a GANs & Increase in average rate of  transmission for UAV downlink communications by over 10\%\\ \hline
    \cite{2-mcenroe2022survey} & 2022  & \cmark & Edge AI (Transfer learning and knowledge distillation, optimizing AI model inference speed)& \cmark &  Optimizing AI model inference speed through pruning and distillation, and integrating reinforcement learning for dynamic data and network management & UAV based IoT services and  identified implementation challenges \\ \hline
       \cite{41-9771337} & 2022 &\cmark & Multi agent deep deterministic technology &\cmark &AI-Based UAV optimization & 3C resource allocation for maximizing energy efficiency and minimizing latency are discussed, it did not explore  the performance of other parameters such as data rate and power \\ \hline
          \cite{14-yang2022generative} & 2022  & \cmark  & Generative AI GANs & \cmark & Significance of trust in the emerging 6G wireless networks and emphasize the role of AI in achieving trustworthy services & Quality of service requirements for UAV-assisted networks can be met\\ \hline
 \cite{1-bilgram2023accelerating} & 2023 & \cmark & Generative AI LLMs& \xmark & Digital Prototyping is used which have single design to bridge the gaps between different entities of system & Can be used by nontechnical persons and management of UAV-assisted networks \\ \hline

  \cite{39-karapantelakis2023generative} & 2023  &\cmark & Generative AI in general &\cmark & Challenges such as  data observability and safe learning are discussed with use cases &  Customer incident management, network planning and deployment are discussed in detail\\ \hline
   \cite{10-zou2023wireless} & 2023  & \cmark &  Generative LLMs& \cmark & On-device Large Language Models (LLMs) and their collaborative role in achieving network goals, emphasizing the limitations of cloud-based LLMs & Multi-agent systems in wireless networks is pertinent to generative AI and UAV-assisted networks, the collaborative potential for efficient and adaptive communication structures in dynamic environments \\ \hline
\cite{11-raha2023generative} & 2023  & \cmark &Generative AI GANs& \cmark & Novel semantic communication framework for next-generation wireless networks to address the challenges of excessive data transmission, high bandwidth consumption, increased latency, and reduced quality of services for applications (e.g., intelligent transportation systems &  UAV-assisted networks are also part of future wireless networks so that it will be a useful work in the broader domain of UAV-assisted IoT networks \\ \hline
 \cite{12-du2023generative} & 2023  & \cmark & AI-Generated Content (AIGC) models& \cmark & Deploying AI-Generated Content (AIGC) models for the emerging Metaverse paradigm by introducing an AIGC-as-a-service architecture in wireless edge networks & Can be useful in UAV-assisted networks where environmental uncertainty and variability are a great challenge \\ \hline
  \cite{13-wang2023unified} & 2023  & \cmark & AI-Generated Content (AIGC) models & \cmark &  Novel framework that utilizes wireless perception to guide generative AI for digital content generation services in resource-constrained mobile edge networks & Proposed a pricing strategy (diffusion model generated optimal pricing strategy) for UAV-assisted networks using channel state information\\ \hline
\end{tabular}
\end{table*}

\begin{figure*} [t]
\centerline{
\includegraphics[width=6.5in, trim = 0.5cm 0.5cm 0.5cm 0.5cm,clip]{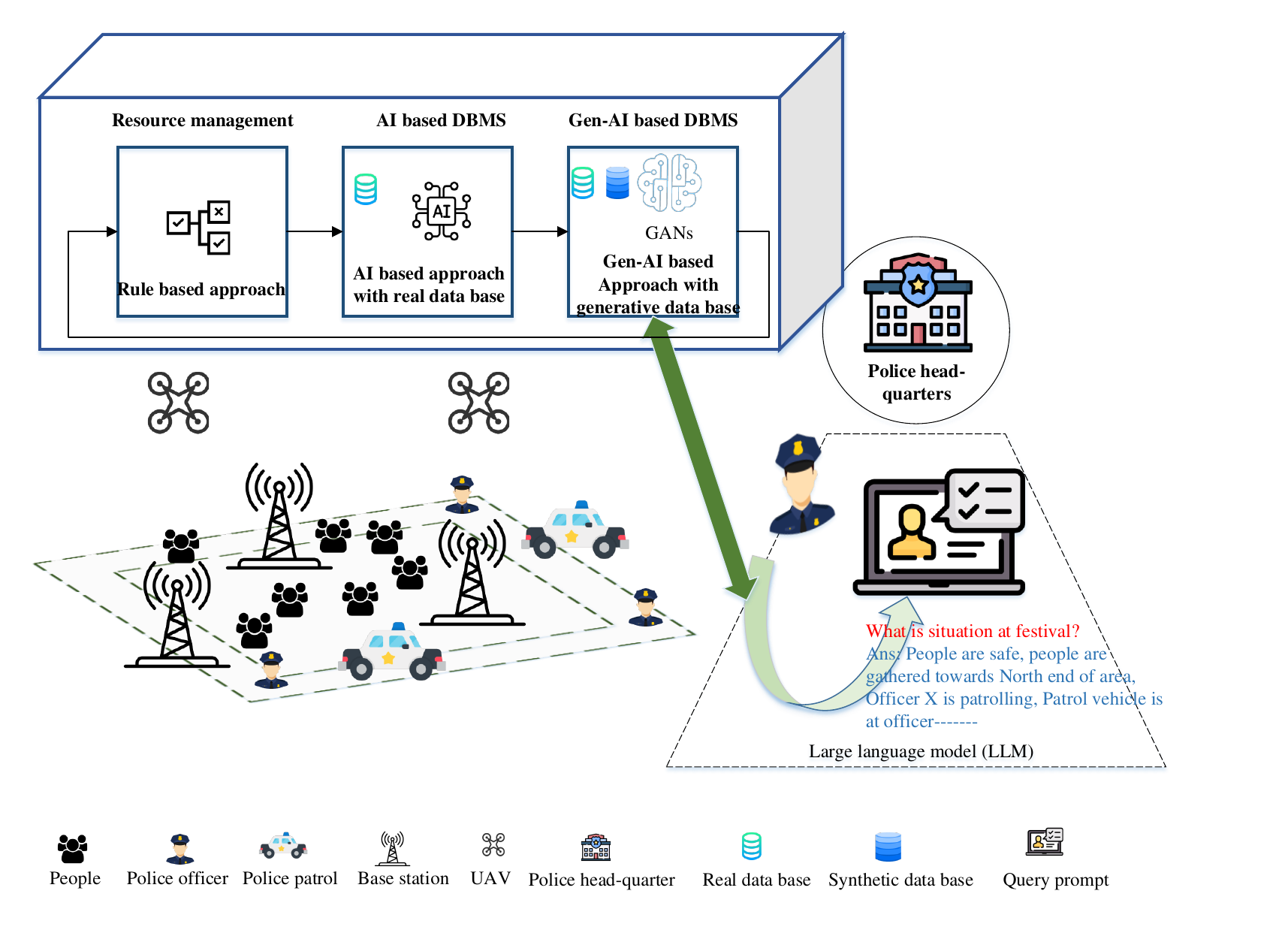}
}
\caption{Generative AI for UAV-assisted IoT networks in public safety.}
\label{fig:system}
\end{figure*}

\section{Use Case: Public Safety}
In this use case, law enforcement authorities have deployed an advanced radio access network, including fixed and mobile UAV base stations, to ensure public safety at a festive event. The police officials at the headquarters monitor the dynamic atmosphere using a state-of-the-art LLM system. The police officers on foot and patrol vehicles navigate the event grounds to maintain public safety. The UAV network provides a comprehensive aerial view, which helps manage resources in a dynamic environment. The LLM's model at the police headquarters serves as the operational center, allowing officers to query the system in real time for situational insights. The generative AI provides valuable information about crowd concentrations and the precise locations of patrolling officers and vehicles, enabling a coordinated and proactive approach to security management. Integrating UAVs, patrol officers, and generative AI ensures the safety of event attendees. Figure \ref{fig:system} illustrates the overall system with the generative AI and UAV-assisted network working in harmony.

Here, we explore various resource management strategies used in our case study. These strategies include rule-based optimization, AI-powered resource allocation, and the integration of GAN and LLM in the resource allocation block. By adding each block to the system, the system achieves more efficient and real-time resource allocation by incorporating these strategies. Managing resources using a traditional algorithmic-based approach can pose significant challenges, especially when ensuring seamless handovers between ground base stations. Estimating user capacity, positions, and available base stations in real-time can be difficult in dynamic scenarios, such as those involving mobile UAVs. An alternative approach is to use a machine learning-based agent to predict optimal handovers. However, this requires effective resource management procedures, especially in complex scenarios.

\subsection{Rule-Based Optimization: Traditional Strategies for Resource Management}
The first approach for resource management is based on predefined rules and strategies to allocate resources efficiently. This method uses traditional optimization principles (e.g., greedy algorithm, genetic algorithm, or dynamic programming algorithms) to handle the complexities of resource allocation in changing environments. Rule-based optimization guarantees efficient resource allocation based on established norms regarding public safety. This approach adjusts dynamically to the real-world situation, enabling quick decision-making for priority users such as police officers. The result is a streamlined and responsive resource management system that improves operational efficiency during public events. 

\subsection{AI-Powered Resource Allocation: Leveraging Real-Time Data for Public Safety}
The second approach proposes an AI-based resource management system (e.g., reinforcement learning, deep Q-networks, deep reinforcement learning, and swarm intelligence algorithms) that uses real-time data from the database management system. This system is an innovative solution that enables dynamic and adaptive resource allocation by utilizing the power of AI to provide intelligent insights and solutions based on real-world scenarios. The AI-powered resource allocation system in public safety uses real-time data from the database management system, which offers dynamic insights crucial for adaptive resource deployment. This integration allows the system to respond promptly to evolving situations, optimizing resource utilization based on continuous data influx. As a result, public safety operations benefit from improved decision-making and enhanced responsiveness to changing conditions.

\subsection{GAN and LLM Integration: Revolutionizing Resource Management in Public Safety}
There is a cutting-edge approach that combines GANs and LLMs (e.g., deep generative models, conditional generative models) into a database management system. This technique optimizes resource allocation dynamically by utilizing the capabilities of GANs and LLMs. Integrating advanced technologies enhances efficiency and ensures adaptability to the ever-changing public safety demands.  GANs enrich the dataset by providing realistic scenarios for training machine learning-based agents. This integration enhances real-time decision-making capabilities, ensuring smooth handovers and quality of service for priority users in a dynamic and challenging environment. The outcome is an adaptive, technologically advanced solution that significantly improves the effectiveness of public safety operations during event management scenarios. This deployment significantly enhances priority users' performance, ensuring service quality during handovers. The result is a comprehensive solution that improves public safety services by enhancing decision-making capabilities, providing realistic scenarios for training, and enabling smooth handovers.

Algorithm \ref{alg:resource_management} presents a step-by-step guide for implementing the functionality of the public safety case study.

\begin{algorithm}
\caption{Public Safety Resource Management}\label{alg:resource_management}
\begin{algorithmic}[1]
\STATE \textbf{INPUT:} Query for event's current situation, e.g., "What is the safety situation at the event/festival?"
\STATE \textbf{Rule-Based Optimization:} 
   \STATE  Determine resources needed based on event size and nature.
   \STATE  Apply priority (police officers) based resource optimization algorithm, e.g., genetic algorithm or dynamic programming algorithm. 
\STATE\textbf{ AI-Powered Resource Allocation:}
   \STATE Collect event information, including location and crowd population and make a database management system. 
   \STATE Apply AI algorithms (such as reinforcement learning, deep Q-networks, deep reinforcement learning, swarm intelligence algorithms) on a continuous influx of data to respond promptly to evolving situations.
\STATE \textbf{GAN and LLM Integration:}
   \STATE Generate scenarios and simulations based on pre-processed data.
   \STATE Predict potential emergencies, crowd behavior, and resource requirements using Generative AI models (such as deep generative models, conditional generative).
\STATE \textbf{OUTPUT:} Response indicating the current situation: "People are safe and have  gathered towards the North end of the area; officer X patrolled the area while patrol vehicles and UAVs patrolled North-South."

\end{algorithmic}
\end{algorithm}


\section{Challenges, Future Research Directions and Recommendations}
The utilization of generative AI in UAV-assisted IoT network resource allocation and real-time requirements poses several challenges that require innovative solutions for practical applications. The following sections discuss these challenges, future research directions, and recommendations.

 \subsection{Optimizing computational efficiency}
The computational complexity of generative models poses several challenges for UAV-assisted IoT networks. These challenges include energy efficiency, real-time requirements, quality of service, and the delicate balance between speed and accuracy. A multi-disciplinary approach is required to overcome these challenges, considering potential solutions such as distributed computing and specialized hardware considerations. By addressing these challenges, we can ensure the successful deployment of generative models in UAV-assisted IoT networks.

\subsection{Adaptable scaling and efficiency}
As the number of connected devices, services, and data types in UAV-assisted IoT networks is expected to grow exponentially, the need of generative models that can meet high engineering requirements for scalability and flexibility will also increase. Horizontal scaling, resource efficiency, adaptive architectures, and data stream adaptability are crucial elements in handling increased computational demands and diverse data scenarios. Decentralization, edge computing, and modular design principles are recommended to enhance scalability and flexibility effectively.

\subsection{Enhancing robustness and reliability}
For generative models to be effective in UAV-assisted networks in IoT systems, the system must be able to handle errors and variations in different environmental and contextual settings. These challenges relate to ecological changes, fault tolerance, uncertainty quantification, and data quality. Optimizing software and hardware aspects and combining data from multiple sensors is essential to ensure robustness.

\subsection{Creating a unified ecosystem}
Achieving smooth interoperability among generative models across diverse devices, vendors, and networks in the context of UAV-assisted networks in IoT systems is a complex endeavor. Several challenges are related to the generative model architecture, including variability, legacy systems, middleware, versioning, security, and standardization efforts. We must establish a standardized, interoperable, and backward-compatible ecosystem to address these challenges. This can be achieved by emphasizing the coordination required from regulatory bodies and industry stakeholders.

\subsection{Navigating regulatory landscapes}
As UAV-assisted IoT networks become more common, generative models become increasingly important. However, several regulatory challenges and policy considerations must be addressed. These issues include data protection, ethics, social impact, national security, cross-border coordination, public/private partnerships, and the role of standardization bodies. It is crucial to ensure compliance without compromising functionality, and ethical guidelines and a multi-stakeholder approach should be emphasized to address these regulatory issues effectively.

\section{Conclusion}
Generative AI in UAV-assisted IoT networks has the potential to revolutionize UAV utilization. The use of generative AI in UAV-assisted IoT networks, such as in a case involving the deployment of law enforcement authorities at a festive event, demonstrates the transformative potential of this technology in ensuring public safety. These approaches enable dynamic and adaptive resource allocation, facilitating a coordinated and proactive approach to security management. In crucial areas like emergency response and environmental monitoring, where resource efficiency is paramount, generative AI offers versatile capabilities to address challenges in resource allocation and adaptability. Generative AI enhances training datasets, as seen in public safety applications, and ensures more efficient and responsive UAV-assisted IoT networks. 
\section*{Acknowledgment}
We thank the anonymous reviewers for their valuable comments which helped us improve the content, organization, and presentation of this paper.
\bibliography{references} 

\begin{thebibliography}{10}

\bibitem{2-mcenroe2022survey}
P.~McEnroe, S.~Wang, and M.~Liyanage, ``{A survey on the convergence of edge computing and AI for UAVs: Opportunities and challenges},'' {\em IEEE Internet of Things Journal}, vol.~9, no.~17, pp.~15435--15459, May 2022.

\bibitem{40-9501056}
Q.~Zhang, A.~Ferdowsi, and W.~Saad, ``{Distributed generative adversarial networks for mmWave channel modeling in wireless UAV networks},'' in {\em IEEE International Conference on Communications (IEEE ICC)}, pp.~1--6, Jun. 2021.

\bibitem{13-wang2023unified}
J.~Wang, H.~Du, D.~Niyato, J.~Kang, Z.~Xiong, D.~Rajan, S.~Mao, {\em et~al.}, ``{A unified framework for guiding Generative AI with wireless perception in resource constrained mobile edge networks},'' {\em IEEE Transactions on Mobile Computing}, pp.~1--17, Mar. 2024.

\bibitem{22-bond2021deep}
S.~Bond-Taylor, A.~Leach, Y.~Long, and C.~G. Willcocks, ``{Deep generative modelling: A comparative review of VAEs, GANs, normalizing flows, energy-based and autoregressive models},'' {\em IEEE Transactions on Pattern Analysis and Machine Intelligence}, vol.~44, no.~11, pp.~7327--7347, Nov. 2022.

\bibitem{24-xu2023unleashing}
M.~Xu, H.~Du, D.~Niyato, J.~Kang, Z.~Xiong, S.~Mao, Z.~Han, A.~Jamalipour, D.~I. Kim, V.~Leung, {\em et~al.}, ``{Unleashing the power of edge-cloud generative AI in mobile networks: A survey of AIGC services},'' {\em IEEE Communications Surveys \& Tutorials}, Jan. 2024.

\bibitem{46-kasneci2023chatgpt}
E.~Kasneci, K.~Se{\ss}ler, S.~K{\"u}chemann, M.~Bannert, D.~Dementieva, F.~Fischer, U.~Gasser, G.~Groh, S.~G{\"u}nnemann, E.~H{\"u}llermeier, {\em et~al.}, ``{ChatGPT for good? On opportunities and challenges of large language models for education},'' {\em Learning and individual differences}, vol.~103, p.~102274, Apr. 2023.

\bibitem{1-bilgram2023accelerating}
V.~Bilgram and F.~Laarmann, ``{Accelerating innovation with Generative AI: AI-augmented digital prototyping and innovation methods},'' {\em IEEE Engineering Management Review}, vol.~51, no.~2, pp.~18--25, Jun. 2023.

\bibitem{44-inayat2022learning}
U.~Inayat, M.~F. Zia, S.~Mahmood, H.~M. Khalid, and M.~Benbouzid, ``{Learning-based methods for cyber attacks detection in IoT systems: A survey on methods, analysis, and future prospects},'' {\em Electronics}, vol.~11, no.~9, p.~1502, May 2022.

\bibitem{45-rafique2020communication}
Z.~Rafique, H.~M. Khalid, and S.~Muyeen, ``{Communication systems in distributed generation: A bibliographical review and frameworks},'' {\em IEEE Access}, vol.~8, pp.~207226--207239, Nov. 2020.

\bibitem{39-karapantelakis2023generative}
A.~Karapantelakis, P.~Alizadeh, A.~Alabassi, K.~Dey, and A.~Nikou, ``{Generative AI in mobile networks: A survey},'' {\em Annals of Telecommunications}, vol.~79, pp.~15--33, Feb. 2024.

\bibitem{41-9771337}
S.~Fu, M.~Zhang, M.~Liu, C.~Chen, and F.~R. Yu, ``{Toward energy-efficient UAV-assisted wireless networks using an artificial intelligence approach},'' {\em IEEE Wireless Communications}, vol.~29, no.~5, pp.~77--83, Oct. 2022.

\bibitem{10-zou2023wireless}
H.~Zou, Q.~Zhao, L.~Bariah, M.~Bennis, and M.~Debbah, ``{Wireless multi-agent generative AI: From connected intelligence to collective intelligence},'' {\em arXiv preprint arXiv:2307.02757}, Jul. 2023.

\bibitem{11-raha2023generative}
A.~D. Raha, M.~S. Munir, A.~Adhikary, Y.~Qiao, and C.~S. Hong, ``{Generative AI-driven semantic communication framework for nextG wireless network},'' {\em arXiv preprint arXiv:2310.09021}, Oct. 2023.

\bibitem{12-du2023generative}
H.~Du, Z.~Li, D.~Niyato, J.~Kang, Z.~Xiong, H.~Huang, and S.~Mao, ``{Diffusion-based reinforcement learning for edge-enabled AI-generated content services},'' {\em IEEE Transactions on Mobile Computing}, pp.~1--16, Jan. 2024.

\bibitem{14-yang2022generative}
L.~Yang, Y.~Li, S.~X. Yang, Y.~Lu, T.~Guo, and K.~Yu, ``{Generative adversarial learning for intelligent trust management in 6G wireless networks},'' {\em IEEE Network}, vol.~36, no.~4, pp.~134--140, Oct. 2022.

\end{thebibliography}
\bibliographystyle{ieeetr}

\end{document}